\documentclass[aps,groupedaddress,prl,twocolumn,showpacs,superscriptaddress]{revtex4}

\usepackage{epsfig}
\usepackage{graphicx}
\usepackage{amsmath,amssymb,amsfonts}
\usepackage{natbib}

\begin{document}

\title{Hydrodynamic View of Wave-Packet Interference: Quantum Caves}

\author{Chia-Chun Chou}
\affiliation{Institute for Theoretical Chemistry and Department of
Chemistry and Biochemistry, The University of Texas at Austin, Texas
78712, USA}

\author{\'Angel S. Sanz}
\affiliation{Instituto de F\'{\i}sica Fundamental,
Consejo Superior de Investigaciones Cient\'{\i}ficas,
Serrano 123, 28006 Madrid, Spain}

\author{Salvador Miret-Art\'es}
\affiliation{Instituto de F\'{\i}sica Fundamental,
Consejo Superior de Investigaciones Cient\'{\i}ficas,
Serrano 123, 28006 Madrid, Spain}

\author{Robert E. Wyatt}
\affiliation{Institute for Theoretical Chemistry and Department of
Chemistry and Biochemistry, The University of Texas at Austin, Texas
78712, USA}

\date{\today}

\begin{abstract}
Wave-packet interference is investigated within the complex quantum
Hamilton-Jacobi formalism using a hydrodynamic description. Quantum
interference leads to the formation of the topological structure of
{\it quantum caves} in space-time Argand plots. These caves consist
of the vortical and stagnation tubes originating from the
isosurfaces of the amplitude of the wave function and its first
derivative.
Complex quantum trajectories display counterclockwise helical
wrapping around the stagnation tubes and hyperbolic deflection near
the vortical tubes.
The string of alternating stagnation and vortical tubes is sufficient
to generate divergent trajectories.
Moreover, the average wrapping time for trajectories and the rotational
rate of the nodal line in the complex plane can be used to define the
lifetime for interference features.
\end{abstract}

%\pacs{03.65.Nk, 03.65.Ta, 03.65.Ca}

\maketitle

%%%%%%%%%%%%%%%%%%%%%%%%%%%%%%%%%%%%%%%%%%%%%%%%%%%%%%%%%%%%%%%%%%%%%%%

One of the most fundamental but intriguing microscopic effects is
quantum interference, the observable feature arising from the coherent
superposition of quantum probability amplitudes.
Quantum interference is involved in a very wide range of experiments
arising from myriad applications. Just to mention some of them, there
are superconducting quantum interference devices \cite{scalapino},
coherent control of chemical reactions \cite{paul}, atom and molecular
interferometry \cite{berman} (including Bose-Einstein condensates
\cite{pritchard}), or Talbot/Talbot-Lau interferometry with
relatively heavy particles (e.g., Na atoms \cite{chapman2} and
Bose-Einstein condensates \cite{deng}).
However, despite all this experimental and theoretical work, very
little attention beyond the implications of the superposition principle
has been devoted to understanding quantum interference at a more
fundamental level \cite{angel-jpa}.

In this Letter, we focus on the hydrodynamical interpretation
\cite{madelung,wyatt-bk} of experiments of this type by introducing
complex quantum trajectories originating as characteristics of the
solutions of the complex quantum Hamilton-Jacobi equation
\cite{CQHJE1,CQHJE2,CQHJE3,CQHJE4,CQHJE5,CQHJE6}.
As recently shown \cite{Tannor}, interference effects on the real axis
may be described in terms of the superposition of amplitudes carried by
approximate (low-order) complex quantum trajectories.
Unfolding of the dynamics from real space into the complex plane yields
unexpected and surprising features, including what we term {\it quantum
caves}.
These caves are topological structures developed around curves in
complex coordinate space where the total wave function and its first
derivative are zero ({\it nodes} and {\it stagnation points},
respectively).
Quantum caves are then displayed in 3D Argand plots (the third
dimension being time), where vortical and stagnation tubes form around
nodal and stagnation curves (which arise from the time-evolution of
nodes and stagnation points, respectively), displaying analogies to
the {\it stalactites} and {\it stalagmites} of real geological caves.

The equation of motion for complex quantum trajectories arises after
substituting the complex-valued wave function in the form $\Psi(x,t)=
\exp[iS(x,t)/\hbar]$ into the time-dependent Schr\"odinger equation.
This yields the complex-valued quantum Hamilton-Jacobi equation
\begin{equation}
 - \frac{\partial S}{\partial t} = \frac{1}{2m} \left( \frac{\partial
 S}{\partial x} \right )^2 +V(x)+ \frac{\hbar}{2 m i}  \frac{\partial
 ^2 S}{\partial x^2} ,
 \label{CQHJ}
\end{equation}
where $S(x,t)$ is the complex action and the last term is the complex
quantum potential, $Q(x,t)$.
For the system studied here, no external interaction potential is
assumed (i.e., $V = 0$).
Quantum trajectories are then developed from the guidance condition
$p(x,t) = \partial S(x,t)/\partial x$, which defines the {\it quantum
momentum function} (QMF).
By analytical continuation, the $x$ variable is extended to the
complex plane through the $z=x + i y$ complex variable (time remains
real-valued) and complex quantum trajectories are determined from
$p(z,t)=\partial S(z,t)/\partial z=(\hbar/i) \partial\ln\Psi(z,t)/
\partial z$.
Two kinds of singularities are especially relevant: (i) {\it nodes}
of the wave function, which correspond to \emph{poles} of the QMF,
and (ii) {\it stagnation points} \cite{chia-polya}, which occur
where the QMF is zero and correspond to points where the first
derivative of the wave function is also zero.
In addition, {\it caustics} are related to free wave-packet propagation
\cite{angel}.

To illustrate the formation of vortical and stagnation tubes and
quantum caves, we consider the head-on collision of two
one-dimensional Gaussian wave packets \cite{angel-jpa,angel}.
Despite its simplicity, this analytical problem is a representative
of other more complicated, realistic processes characterized by
interference. This process can be described by the total wave
function, $\Psi (x,t) = \psi_L (x,t) + \psi_R (x,t)$ ($L$/$R$
denotes left/right), which is analytically continued to the complex
plane to give $\Psi(z,t)$. Each partial wave is represented by a
free Gaussian wave packet,
\begin{equation}
 \psi (x,t) = A_t \ \! e^{- (x-x_t)^2/4\sigma_t\sigma_0
  + i p (x - x_t)/\hbar + i E t/\hbar} ,
 \label{int2}
\end{equation}
where, for each component, $A_t = (2\pi\sigma_t^2)^{-1/4}$ and the
complex time-dependent spreading is given by $\sigma_t=\sigma_0
(1+i\hbar t/2m\sigma_0^2)$ with the initial spreading $\sigma_0$.
Due to the free motion, $x_t = x_0 + v t$ ($v = p/m$ is the
propagation velocity) and $E = p^2/2m$. We also consider
the case where the relative propagation velocity is larger than the
wave packet spreading rate, $\hbar/2m\sigma_0$ \cite{angel-jpa}.
From now on, all quantities will be given in atomic units ($\hbar=m=1$).

\begin{figure}
 \includegraphics[width=8cm]{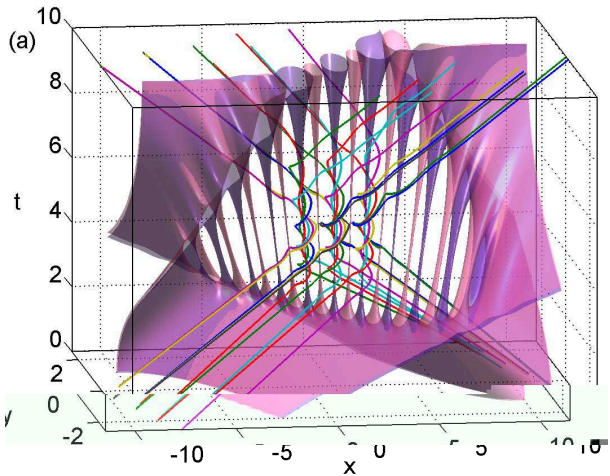}
 \caption{\label{fig1}
  Quantum caves for head-on collision of two Gaussian wave packets.
  These caves are formed with the isosurfaces $|\Psi (z,t)|=0.053$
  (pink/lighter gray sheets) and $|\partial \Psi (z,t)/ \partial z|
  = 0.106$ (violet/darker gray sheets).
  The complex quantum trajectories launched from two branches of
  the isochrone reach the real axis at $t=5$.}
\end{figure}

The following initial conditions are used: $x_{0L} = - 10 = - x_{0R}$,
$v_{L} = 2 = -v_{R}$ and $\sigma_0 = \sqrt{2}$, and maximal interference
occurs at $t=5$ in real space.
In Fig.~\ref{fig1}, complex quantum trajectories together with the
isosurfaces $|\Psi (z,t)| = 0.053$ (pink/lighter gray sheets) and
$|\partial \Psi (z,t)/\partial z|= 0.106$ (violet/darker gray sheets)
from $t=0$ to $t=10$ are shown in a 3D Argand plot.
Around nodes and stagnation points, tubular shapes develop
(pink/lighter gray and violet/darker gray tubes, respectively), which
alternate with each other and whose centers correspond to vortical and
stagnation curves, respectively.
The sharp features and well defined vertical tubes observed in
Fig.~\ref{fig1}, reminiscent of {\it stalactites} and {\it
stalagmites}, lead us to call these plots {\it quantum caves}.
Interference leads to the formation of quantum caves and produces
this topological structure.

\begin{figure*}
 \includegraphics[width=6.5cm]{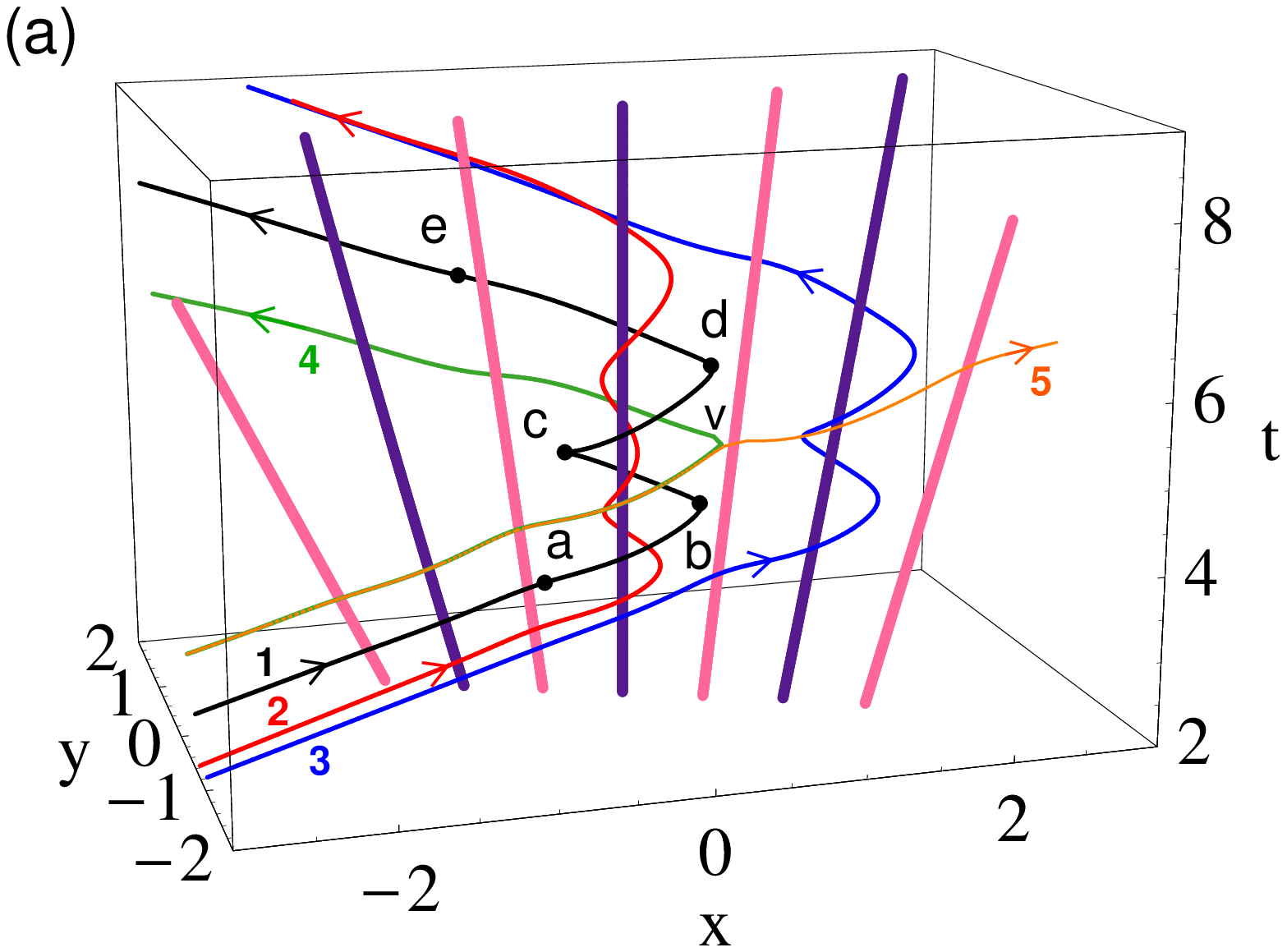}
 \includegraphics[width=6.5cm]{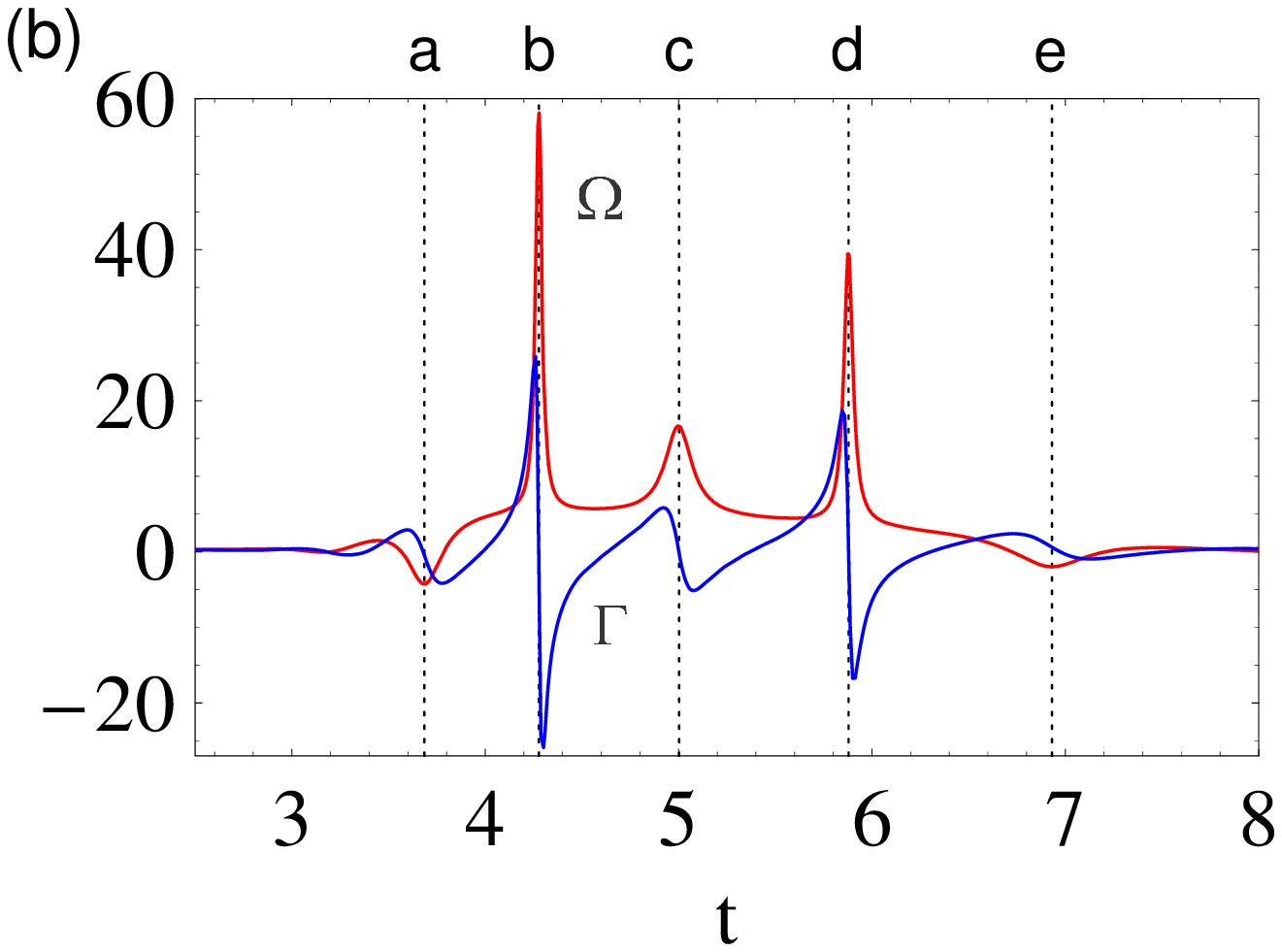}
 \caption{\label{fig2}
  (a) Trajectories 1, 2, 3 and 4 launched from the isochrone which
  arrive on the real axis at $t=5$ display hyperbolic deflection around
  the vortical curves (pink/light gray) and helical wrapping around the
  stagnation curves (violet/dark gray).
  Trajectories 4 and 5 diverge at position ``V'' near the vortical
  curve.
  (b) Divergence and vorticity of the QMF along \mbox{trajectory 1.}}
\end{figure*}

As seen in Fig.~\ref{fig1} and, in more detail, in Fig.~\ref{fig2}(a),
the complex trajectories display counterclockwise helical wrapping
around the stagnation tubes, while they are hyperbolically deflected
or ``repelled'' when they approach the vortical tubes enclosing the
QMF poles.
This intricate motion depicts the probability density flow around the
vortical and stagnation tubes.
Trajectories launched from different initial positions may wrap around
the same stagnation curve and remain trapped for a certain time
interval.
As time proceeds, these trajectories separate from the stagnation
curves in analogy to the decay of a resonant state. Therefore, the
whole process shows long-range correlation among trajectories
arising from different starting points.

The QMF can be viewed as a vector field in the complex plane,
$p=p_x+ip_y$, and we can compute its divergence and vorticity along
a complex quantum trajectory, which describe the \emph{local}
expansion or contraction and rotation of the quantum fluid,
respectively.  By the Cauchy-Riemann equations, the first derivative
of the QMF becomes $\partial p/\partial z=(\Gamma + i\Omega)/2$,
where $\Gamma = \vec{\nabla} \cdot \vec{p} =
\partial p_x/\partial x + \partial p_y/\partial y$ is the divergence
of the QMF and $\Omega = |\vec{\nabla} \times \vec{p}\,| = (\partial
p_y/\partial x - \partial p_x/\partial y)$ is the vorticity.
Moreover, the complex quantum potential in Eq.~(\ref{CQHJ}) can be
expressed in terms of divergence and vorticity by
\begin{equation}
Q(z,t)=\frac{\hbar}{2mi}\frac{\partial p}{\partial
z}=\frac{\hbar}{4mi}\left(\Gamma+i\Omega\right). \label{CQP}
\end{equation}

Figure~\ref{fig2}(a) shows trajectories 1, 2, 3 and 4 launched from
the isochrone which arrive on the real axis at $t=5$ (maximal
interference), and Fig.~\ref{fig2}(b) presents the time evolution of
the divergence and vorticity of the QMF along trajectory 1.  When
the particle approaches the vortical curve at position $a$, it
experiences a repulsive force provided by the pole of the QMF and
the trajectory displays hyperbolic deflection.  As shown in
Fig.~\ref{fig2}(b), $\Gamma$ and $\Omega$ display the first sudden
spike.  Then, this particle is trapped by the stagnation curve
between two vortical curves.  When the trajectory approaches turning
points ($b$, $c$ and $d$), the particle's velocity undergoes rapid
changes and this produces sharp fluctuations in $\Gamma$ and
$\Omega$.  From Eq.~(\ref{CQP}), the quantum potential is larger near
these positions.  Finally, as the particle departs from the
stagnation curve, it experiences a repulsive force provided by the
pole and the trajectory displays hyperbolic deflection at position
$e$.  The whole process indicates important dynamical activity,
which is lacking within the real-valued version of this problem
(where no divergence or vorticity can be defined).

The wrapping time for a specific trajectory can be defined by the
interval between the first and last minimum of $\Omega$, and the
positive vorticity within this time interval describes the
counterclockwise twist of the trajectory.
The sign of $\Gamma$ indicates the local expansion or contraction of
the quantum fluid when it approaches or leaves a turning point,
respectively.
Within this time interval, the particle obviously feels the presence
of stagnation points and nodes, and the trajectory displays the
interference dynamics.  From Fig.~\ref{fig2}(b), the wrapping process
lasts from $t \approx 3.7$ to $t \approx 6.9$.
In addition, trajectories 1 and 2 wrap around the same stagnation curve
with different wrapping times and numbers of loops.
The wrapping time around a stagnation curve is determined by $\Gamma$
and $\Omega$, which are used to characterize the turbulent flow.
Thus, the average wrapping time for those trajectories reaching the
real axis at the time of maximal interference can be used to define
the ``life-time'' for the interference process observed on the real
axis.

Trajectories 2 and 3 start from the isochrone with the initial
separation $\Delta z_0\approx0.3$, wrap around different stagnation
curves, and then end with the separation at $t=10$ $\Delta
z\approx0.8$. These two trajectories avoid the vortical curve and
this greatly increases the separation between them.  This behavior
is consistent with what one observes when looking at the quantum
flow in real space: the trajectory distribution is sparse near nodes
of the wave function and dense between two consecutive nodes.  In
addition, trajectories 4 and 5 start with slightly different initial
positions, $\Delta z_0=0.01$, and they suddenly separate at position
``V'' near the vortical curve.
This leads to the continuously increasing separation between them and
a positive Lyapunov exponent, analogous to the case reported in real
space \cite{CQHJE4}.
The alternating structure for the vortical and stagnation tubes
(similar to that for the nodal point--X-point complex in Bohmian
mechanics in 2D real space) thus leads to divergent trajectories and
may generate chaos \cite{Efthymiopoulos}.

Time-dependent nodal positions in the complex plane can be
determined analytically by solving the equation $\Psi(z,t)=0$, which
renders
\begin{equation}
 z_n(t) = \frac{i\pi\left(n+1/2\right)}
  {\left[imv/\hbar-\left(x_0-vt\right)/\left(2\sigma_0\sigma_t\right)
    \right]} ,
\end{equation}
where $n = 0, \pm1, \pm2, \ldots$  Splitting this expression into
its real and imaginary parts, $z_n(t)=x_n(t)+iy_n(t)$, yields the
analytical expression for the angle of the nodal line with the
positive real axis, $\theta(t)=\tan^{-1}\left[y_n(t)/x_n(t)\right]$,
(which is independent of $n$), and this describes the time evolution
of the nodal line. In addition, the time evolution of the $n$th node
is given by $y_n = (2mv\sigma_0^2/\hbar x_0)x_n -
(2n+1)(\pi\sigma_0^2/x_0)$.

\begin{figure}[b]
 \includegraphics[width=8cm]{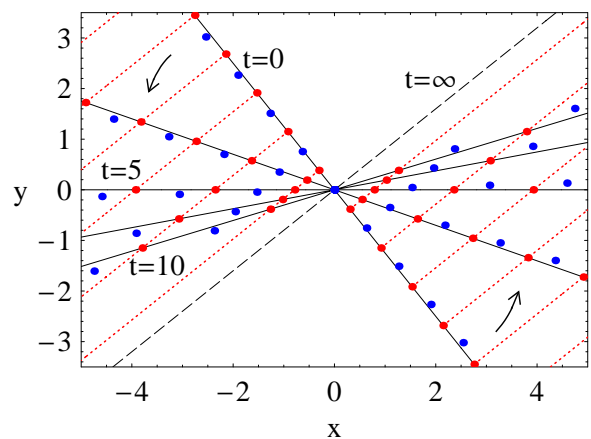}
 \caption{\label{fig3}
  Evolution of the nodal line at $t = 0, 2.5, 5, 7.5, 10$
  (black solid lines) and $t = \infty$ (black dashed line); the arrows
  indicate the rotation direction.
  Nodes and stagnation points are denoted by dots; nodal trajectories
  are shown as dotted lines passing through the nodal points.}
\end{figure}

Figure \ref{fig3} shows the time-dependent string of stagnation
points and nodes and nodal trajectories in the complex plane. At
$t=0$, these two wave packets are far away from each other; however,
their tails interfere in the complex plane. This contributes to the
string of stagnation points and nodes, and the initial angle of the
nodal line (which is perpendicular to nodal trajectories) is
$\theta_0 = \tan^{-1}(-\hbar x_0/2mv\sigma_0^2) = -51.34^\circ$.
Then, the nodal line rotates counterclockwise and crosses the real
axis at $t=5$, where the total wave function displays maximal
interference. At this time, $y_n = 0$ and we recover the expression
for the positions of nodes, $x_n = (n+1/2)\lambda/2$ ($\lambda =
h/mv$). After $t=5$, these two wave packets start to separate, and
the nodal line continues to rotate counterclockwise away from the
real axis. However, these two wave packets still interfere with each
other in the complex plane. When $t$ tends to infinity, the angle of
the nodal line approaches $\theta_\infty =
\tan^{-1}(2mv\sigma_0^2/\hbar x_0)=38.66^\circ$ and this line
becomes parallel to nodal trajectories.  The nodal line rotates
counterclockwise \cite{angel} from $\theta_0$ to a limiting value
$\theta_\infty$ with an angular displacement $\Delta \theta =
\theta_\infty - \theta_0 = \pi/2$, and its intersections with nodal
trajectories determine the positions of nodes.  In addition, the
distance between stagnation points and nodes increases with time. In
particular, if both wave packets are initially very far apart ($x_0
\to \infty$) but move with a finite velocity $v$, or they are
separated by an arbitrary finite distance with $v = 0$, then the
nodal line will end up aligned with the real axis.
Interference features are observed on the real axis only when the
nodal line is near the real axis.
For the case shown in Fig.~\ref{fig3}, this occurs
between about $\theta(3.52) = -10^\circ$ and $\theta(7.32)=
10^\circ$, so that the ``lifetime'' for the interference features
is about $\Delta t=3.8$. Therefore, the lifetime of interference
features observed on the real axis is determined by the rotation
rate of the nodal line in the complex plane, and this rate
$d \theta(t) / dt$ decays monotonically to zero as $t \to \infty$.

The complex quantum trajectory method provides an insightful
alternative to the traditional analysis of quantum interference
phenomena in real space. In Bohmian mechanics, when two or more real
coordinates are involved, quantum vortices form around nodes in the
wave function and streamlines surrounding the vortex core form
approximately circular loops \cite{Hirschfelder,sanz-jcp}. The
interference of two wave packets in one real coordinate leads to the
formation of nodal structure, but quantum trajectories close to
nodes forming at the maximal interference time do \emph{not} display
vortical dynamics \cite{angel}.  The quantum potential near these
nodes forces these trajectories to avoid these regions and to
exhibit laminar flow in space-time plots.  In contrast, complex
quantum trajectories displaying helical wrapping and hyperbolic
deflection undergo turbulent flow in the complex plane.  This
counterclockwise circulation of trajectories launched from different
positions around the same stagnation tubes can be viewed as a
resonance process in the sense that during interference some
trajectories keep circulating around the tubes for finite times and
then escape as time progresses.  On the other hand, in conventional
quantum mechanics, the interference pattern transiently observed on
the real axis is attributed to constructive and destructive
interference between components of the total wave function.  In
contrast, within the complex quantum trajectory formalism, two
counter-propagating wave packets are \emph{always} interfering with
each other in the complex plane.  This leads to a persistent pattern
of nodes and stagnation points which is a signature of the ``quantum
coherence'' demonstrating the connection between both wave packets
before or after interference fringes are observed on the real axis.
The interference features observed on the real axis are connected to
the rotational dynamics of the nodal line in the complex plane.
Therefore, the average wrapping time for trajectories and the
rotation rate of the nodal line in the complex plane provide two
methods to define the \emph{interference lifetime} observed on the
real axis.  This analysis demonstrates that the complex quantum
trajectory method provides a novel perspective and leads to new
insights for analyzing and interpreting quantum mechanical problems.
Finally, similar conclusions are drawn when the spreading velocity
of the wave packets is greater than their propagation velocity.

%%%%%%%%%%%%%%%%%%%%%%%%%%%%%%%%%%%%%%%%%%%%%%%%%%%%%%%%%%%%%%%%%%%%%%%

C.-C. Chou and R.\ E.\ Wyatt thank the Robert Welch Foundation for the
financial support of this research; A.S.\ Sanz and S.\ Miret-Art\'es
acknowledge the Ministerio de Ciencia e Innovaci\'on (Spain) for
financial support under Project FIS2007-62006.
A.\ S.\ Sanz also acknowledges the Consejo Superior de Investigaciones
Cient\'{\i}ficas for a JAE-Doc contract.

%%%%%%%%%%%%%%%%%%%%%%%%%%%%%%%%%%%%%%%%%%%%%%%%%%%%%%%%%%%%%%%%%%%%%%%

\end{document}